*Research Article*

# Efficient Byzantine Consensus Mechanism Based on Reputation in IoT Blockchain

**Xu Yuan** ,[1] **Fang Luo** ,[1] **Muhammad Zeeshan Haider** ,[1] **Zhikui Chen** ,[1] and **Yucheng Li** [2]

[1]*School of Software, Dalian University of Technology, Dalian 116620, China*
[2]*University of California, Berkeley CA 94720, USA*

Correspondence should be addressed to Xu Yuan; david@dlut.edu.cn

Blockchain technology has advanced rapidly in recent years and is now widely used in a variety of fields. Blockchain appears to be one of the best solutions for managing massive heterogeneous devices while achieving advanced data security and data reputation, particularly in the field of large-scale IoT (Internet of Things) networks. Despite the numerous advantages, there are still challenges while deploying IoT applications on blockchain systems due to the limited storage, power, and computing capability of IoT devices, and some of these problems are caused by the consensus algorithm, which plays a significant role in blockchain systems by ensuring overall system reliability and robustness. Nonetheless, most existing consensus algorithms are prone to poor node reliability, low transaction per second (TPS) rates, and scalability issues. Aiming at some critical problems in the existing consensus algorithms, this paper proposes the Efficient Byzantine Reputation-based Consensus (EBRC) mechanism to resolve the issues raised above. In comparison to traditional algorithms, we reinvented ways to evaluate node reliability and robustness and manage active nodes. Our experiments show that the EBRC algorithm has lower consensus delay, higher throughput, improved security, and lower verification costs. It offers new reference ideas for solving the Internet of Things+blockchain+Internet court construction problem.

## 1. Introduction

The Internet of Things (IoT) is a loosely coupled system of various heterogeneous and homogeneous devices with sensing, processing, and networking capabilities [1, 2]. In recent years, IoT technology has increasingly matured and been applied to a wide range of industries, including creating an Internet court that employs modern tools or technology to create a highly intelligent process and management. However, due to the rapid growth of IoT technology, some indiscreet applications expose sensitive data and user privacy to security threats such as DDoS and Sybil attacks [3], as well as system failures.

Recently, applying blockchain technology in IoT has attracted widespread attention in both industry and academia[4–6]. As a unique ledger technology, blockchain has the characteristics of decentralized, nontampering, programmable, and encrypted [7–9]. However, due to the insufficient storage capacity, low computing and processing power of the Internet of Things blockchain, insufficient data transmission bandwidth, etc., the throughput of the IoT blockchain needs to be optimized [10, 11]. The consensus mechanism is the most crucial factor affecting IoT blockchain throughput. The consensus mechanism, which is the core of the blockchain system, significantly impacts the network's throughput, delay, and fault tolerance. Furthermore, due to the emergence of 5G mobile communications and the exponential growth of IoT devices, reliable high-throughput performance needs to consider the fundamental attribute of the future blockchain architecture to handle many massive data transactions from the IoT network. Therefore, the scalability of the consensus algorithm is a fundamental feature of the blockchain system that must be improved in the future [12].

As the core of a blockchain system, consensus algorithms significantly impact the overall throughput, delay, and fault tolerance of the blockchain network. Typical consensus algorithms can be divided into proof-based consensus and voting-based consensus [13, 14]. Proof-based consensus algorithms include PoW (Proof of Work) [15–17], PoS (Proof of Stake) [18–20], and DPoS (Delegated Proof of Stake) [21]. This algorithm requires nodes joining the network to prove that they are more qualified than other nodes to add a block to the chain. Consensus algorithms based on voting include PBFT (Practical Byzantine Fault Tolerance) [22, 23] and Terdermint [24]. This algorithm requires network nodes to exchange the most recent new block or transaction verification results. Besides, there are many mixed consensuses [25–28]. Their core idea is to learn from each other's strengths and improve the efficiency and robustness of the original consensus.

Algorithms for reaching a consensus must find the right balance between fairness and computational efficiency, and various algorithms have their own advantages and disadvantages. However, considering the limited abilities of IoT blockchain nodes, the PBFT consensus protocol is recognized as the best protocol in the current IoT blockchain system since the PBFT algorithm can avoid the fact that the proof-based consensus algorithm consumes a significant amount of computational resources (such as PoW). Furthermore, each node does not need to verify and store each block [15, 17], while allowing less than $(n-1)/3$ nodes to fail for various reasons, the system continues to run normally, and nodes become malfunctioned for various reasons. However, the PBFT consensus algorithm still have some problems [29]: (1) the PBFT algorithm does not evaluate the reliability of the nodes, resulting in unreliable nodes participating in the consensus; (2) the PBFT algorithm does not substantially penalize malicious nodes that will cause malicious nodes repeatedly attacking the system; (3) the PBFT algorithm works well on small fixed-scale networks where nodes may freely join or leave, which may not be sufficient for large-scale IoT networks with a large number of complex nodes; and (4) the PBFT protocol generates high traffic overhead due to frequent network communication $n^2$ between $n$ nodes.

Therefore, a large number of solutions to the above problems have been proposed. For example, the blockchain community NEO proposed the Delegated-BFT (DBFT) consensus algorithm [30]. In DBFT, nodes use their tokens as votes to select a group of consensus nodes to perform PBFT. However, the election method in DBFT only relies on tokens, which increases the capacity of nodes with more tokens, thereby weakening the degree of decentralization of the system. Li et al. proposed the EPBFT consensus algorithm [31], which relies on VRF (verifiable random function) to screen out consensus nodes for consensus, which improves the security of the blockchain system. However, it did not make a detailed assessment of the reliability of the nodes. Trust-PBFT [32] and T-PBFT [33] try to combine trust with the PBFT algorithm and select the appropriate number of participants from an extensive blockchain network through trust. However, this type of algorithm does not solve the communication of PBFT high complexity issues. Yuanchao et al. [34] evaluated the reliability of nodes according to the node division mechanism and dynamically selected nodes with high reputation values to participate in the consensus and increase the security of the system. However, the detailed reputation calculation scheme was not given in the article.

Although the above consensus algorithm optimizes PBFT from various angles, compared with traditional PBFT, it does improve the consensus efficiency to a certain extent. However, they are not based on the reliability of nodes to screen consensus nodes and cannot fundamentally overcome the shortcomings of traditional PBFT. Therefore, to solve the above problems, we propose an efficient consensus mechanism (EBRC) based on Byzantine reputation, which supports large-scale and dynamic IoT blockchain systems. The overall model is divided into five parts, as shown in Figure 1. In general, our contributions are as follows.

(i) We propose a novel reputation-based blockchain consensus protocol EBRC for IoT-blockchain applications. The protocol utilizes reputation information and timestamp from IoT devices to ensure the reliability of nodes and enhance blockchain security by a deposit penalty mechanism. The reputation value is closely related to the node's behavior. It can be used as an incentive mechanism to enable IoT devices to behave well to become consensus nodes

(ii) VRF random election mechanism in EBRC ensures fairness and randomness when electing nodes in the IoT blockchain system. Meanwhile, comparing with the classic PBFT, the consensus protocol EBRC has comprehensively improved the consensus network's scalability in the IoT blockchain and reduced the communication complexity between nodes

(iii) EBRC also features a dynamic joining and exiting protocol (DJEP), lacking in PBFT. With this mechanism, blockchain systems can quickly adapt to new network status and reach consensus

(iv) We build a blockchain prototype with the EBRC protocol and conducted extensive experiments with around 40 nodes (simulate server machines as IoT devices). The proposed EBRC algorithm shows existing algorithms like PBFT with higher consensus efficiency, throughput, and lower communication overhead in the network

The rest of this paper is organized as follows. Section 2 will systematically introduce the EBRC protocol and algorithm; then, we will describe the realization and evaluation of the proposed consensus protocol in Section 3. Finally, we summarize this paper in Section 4.

## 2. The Framework of the Agreement

This section will elaborate on the design of the EBRC protocol in detail with an overview following by stepwise fundamental component analysis.

*2.1. Protocol Overview.* The overall process framework of the reputation consensus protocol EBRC applicable to the IoT blockchain is shown in Figure 2. The consensus protocol EBRC comprises reputation evaluation, reputation growth rate evaluation, VRF random election mechanism, EBRC consensus protocol, and dynamic join and exit protocol DJEP.

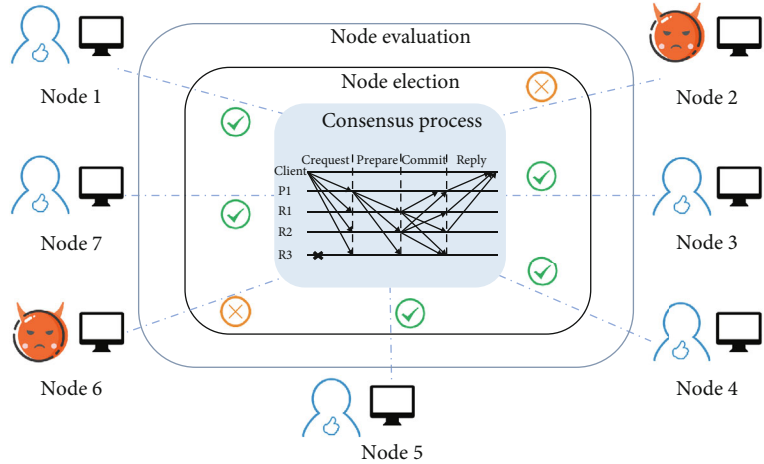

Figure 1: In the proposed EBRC consensus protocol, Node 1, Node 3, Node 4, Node 5, and Node 7 in the figure are normal nodes, while Node 2 and Node 6 are malicious nodes. After the nodes are first evaluated, normal nodes can enter the node election link, and after the election, they become consensus nodes for consensus.

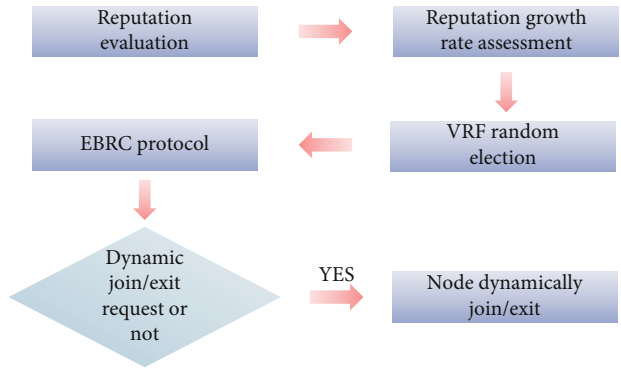

Figure 2: Overall model framework. The EBRC is made up of five components.

Firstly, To ensure the high-quality consensus node collection, the reputation evaluation algorithm evaluates the reputation and reputation growth rate based on the nodes' various indicators in the block node-set composed of all nodes. Secondly, applying the VRF function, we randomly select nodes with the election authority to ensure the randomness and fairness of the election. Thirdly, according to the proposed EBRC consensus protocol, the consensus' validity and scalability are ensured. Finally, according to the DJEP protocol, the dynamic joining and exiting of nodes in the system are completed, enhancing the dynamics of the blockchain system. In general, the reputation-based consensus protocol EBRC effectively improves the reliability of nodes and system security and ultimately enhances the scalability of the blockchain network.

*2.2. Node Evaluation.* The network in the EBRC protocol is composed of $N$ public nodes; each of them has a reputation value and a reputation value growth rate, which determines whether the node can become a consensus node. Further-more, we assume that the network is partially synchronized, meaning that the transaction is provided to the network by the customer in a prespecified format. The network also runs in epochs as a major cycle, and each epoch is divided into 20 rounds. In each round, one block is generated.

For ease of explanation and generality, we also make the following explanations:

(i) Node election authority: according to the proposed authority table, nodes that meet the election conditions

(ii) Consensus node-set: a set of nodes that have voting rights and are selected through the VRF random function

(iii) Consensus nodes: the representative set of nodes participating in the blockchain consensus divided proportionally from the consensus node-set

(iv) Candidate consensus node: a set of nodes that are quickly replaced when the consensus node fails or when the node exits

(v) Behavior record table: a table that records node reputation, reputation growth rate, and reputation reference factors jointly maintained by all nodes in the entire network, as shown in Table 1. Each node will create a local metadata pool to cache all nodes' behavior records in a table. The table is updated after each epoch

*2.2.1. Reputation Model.* The reputation model can provide a good foundation for the reliability of nodes. On the one hand, dynamically updating the node's reputation value continually reduces the occurrence rate of error nodes in the consensus process, ensuring that the blockchain mechanism is highly accessible and secure. On the other hand, the consensus node's reputation value can be used as a standard of reward and punishment to encourage the node to follow the protocol.

TABLE 1: Behavior record table.

| Attributes | Mark | Attributes | Mark |
| --- | --- | --- | --- |
| Node id | *** | Deposit | $d_i$ |
| Node public key | PK | Total number of consensus participation | $M$ |
| Reputation list | $R_i = [r_1, r_2, \cdots, r_j]$ | The number of times an error message was reported | $e_i$ |
| Reputation growth rate list | $Y(t) = [y_1, y_2, \cdots, y_k]$ | Offline time period level | $t_i$ |
| The number of times the consensus was not successfully completed | $m_i$ | Time period level of network delay | $l_i$ |
| Time level of joining the network | $T_i$ | List of processed transaction sizes | $Z_i = [c_1, c_2, \cdots, c_j]$ |

For each node in the system, an actual number with a reputation value $R$ between (0,1] is defined. The reputation value, which represents the credibility of the corresponding node, is set to 0.5 for newly added nodes. Furthermore, in order to become a consensus node, the node must pay a deposit, which affects the calculation of the reputation value. The reputation evaluation factors are as follows.

*Margin ratio*: the margin of node $i$ is expressed as $d_i$, and the definition $D = \sum d_i$ is the sum of margin in the blockchain network, and the margin ratio is $\delta_i = d_i/D \in (0, 1]$. The security deposit setting penalizes malicious behavior with significant monetary penalties, which reduces the risk of malicious nodes causing damage to the network. If a node pays more margin, it increases its chances of participating in the consensus, but it also increases its penalty for bad conduct. Besides, margin thresholds are set to prevent certain nodes from dominating node nominations by earning large sums of money.

*Incomplete rate*: the ratio of the number of times $m_i$ that the node $i$ has failed to complete the consensus to the total number of times the node participates in the consensus $M$, namely, $\tau_i = m_i/M \in (0, 1]$. The unfinished rate mainly measures the completion of the consensus of the nodes. The key factor here is the number of times the node has failed to complete the consensus due to failure. Through this indicator, nodes with more consensus completion times can be selected first, and the status of these nodes is relatively more stable, thereby increasing the overall consensus completion rate.

*Evil rate*: the ratio of the number of times $e_i$ that node $i$ sends an error message and is successfully reported to the total number of times $M$ that node participates in the consensus, i.e., $\Psi_i = e_i/M \in (0, 1]$. The evil rate mainly determines whether a node sends messages honestly. When a node sends an error message, other nodes will report it, and the report can be counted into the system until a consensus agreement is reached.

*Activity rate*: the time period level of node $i$ offline is recorded as $t_i$, the time period level of node $i$ network latency is recorded as $l_i$, the time period level of node $i$ joining the network is recorded as $T_i$, and the classification of each time period is shown in Table 2. The node activity rate is: $\varphi_i = (t_i + t_i)/T_i \in (0, 1]$. The activity rate mainly measures the activity performance of the node. Nodes in the Byzantine consensus process must broadcast to one another and wait for clarification messages. When low-activity nodes participate in the consensus, it will cause many nodes to wait, causing the consensus to stagnate. As a result of implementing this indicator, nodes with higher activity have a greater chance of being chosen, reducing consensus delay and increasing consensus success rate.

*Transaction magnitude factor*: the transaction magnitude factor is used to identify the historical transaction processing capability of node $i$. Suppose the transaction size list processed by node $i$ is denoted as $Z_i$, and the elements in it are sorted from high to low as $c_1, c_2, \cdots, c_j$, and the transaction magnitude factor $\rho_i$ can be expressed as

$$\rho_i = \begin{cases} j, & j = c_j, \\ j - 1, & j > c_j. \end{cases} \tag{1}$$

The transaction magnitude factor is used to determine node $i$ transaction processing capacity in the past. The design of this indicator is derived from the $H$-index [35], which is also called $H$-factor. The higher the $H$-index of a researcher, the higher his academic status. A node's total processing capacity can be measured by measuring its transaction magnitude factor in the blockchain network. When existing nodes agree on a transaction with a high transaction level, this allows them to achieve higher-quality consensus in order to increase the transaction level factor score.

The degree of control of each parameter on the overall credit value varies throughout the calculation process. This paper focuses on the behavior evaluation of malicious nodes. Therefore, when calculating the weighting factor for the above parameters, we give the most weight to the node's non-completion and evil rates, followed by its operation rate. Finally, the node's margin ratio and transaction magnitude factor have the same weight in the credit value calculation. Based on the above content, the weight of each factor is defined as follows: $\vec{W} = [0.1, 0.3, 0.3, 0.2, 0.1]$. Therefore, the formula for calculating the reputation value $R$ is

$$R_i = 0.1\delta_i + 0.3\tau_i + 0.3\psi_i + 0.2\varphi_i + 0.1\rho_i. \tag{2}$$

TABLE 2: The classification of each time period.

| Offline level $t_i$ | | Network latency level $l_i$ | | Join network level $T_i$ | |
|---|---|---|---|---|---|
| Time (h) | Level | Time (ms) | Level | Time (h) | Level |
| (0,0.5] | 10a | (0, 30] | 10b | (96, +∞) | 10c |
| (0.5,2] | 8a | (30, 50] | 8b | (72, 96] | 9c |
| (2, 24] | 6a | (50, 80] | 6b | (24, 72] | 8c |
| (24, 72] | 4a | (80,100] | 4b | (12, 24] | 4c |
| (72, +∞) | 2a | (100, +∞) | 2b | (12, 24] | 2c |

Based on the above evaluation factors, when a node commits malicious behavior, we can update the node's reputation value in time and perform a penalty mechanism of deposit deduction.

### 2.2.2. Reputation Growth Rate Model.

The reputation growth rate is used to determine the node's robustness. It is possible to assess the node's growth after entering the blockchain network by measuring the reputation growth rate, allowing the node to act reasonably to maintain the reputation growth rate.

We define the reputation growth rate $Y(t)$ as $(0,100\%]$. The reputation value of a node newly added to the blockchain system is initialized to 50%. Based on the node reputation's dynamic changes, its reputation growth rate also changes accordingly. Precisely, the new reputation growth rate of the entire network node is calculated by the following formula:

$$Y(t) = \left[ \left( \frac{R_{i,n}}{R_{i,t}} \right)^{1/(t-1)} - 1 \right] * 100\%. \qquad (3)$$

$R_{i,n}$ represents the reputation value of node $i$ in the current round, and $R_{i,t}$ represents the reputation value of node $i$ in the previous $t$ round. This formula calculates the average value of the growth rate since the node joined the system.

The calculation of reputation value and reputation growth rate is carried out around the behavior record table. Before the end of each epoch, the current consensus master node initiates a request to update the reputation value and reputation growth rate. The specific implementation is shown in Algorithm 1. After receiving the request, the consensus node calculates the new reputation value of the entire network according to the local behavior record buffer pool. It then uses the EBRC consensus protocol proposed in this article to reach a consensus on the EBRC consensus protocol. After the consensus is completed, the master node updates the node's reputation, whose reputation value and reputation growth rate have changed in this round into the behavior record table and broadcast it on the entire network. The system then enters the next round of the consensus election stage.

### 2.3. Node Election.

Nodes with different reputation values and reputation growth rates have different election permissions. To ensure the efficiency and security of the system,

---

**Input:** Node set $N$, Old behavior record table $\mathbb{Z}_{Old}$.
**Output:** New behavior record table $\mathbb{Z}_{New}$
**While** $\mathbb{Z}_{Old}$ **do**
　　Read($\mathbb{Z}_{Old}$);
　　$\delta_i = d_i/D \in (0, 1]$;
　　$\tau_i = m_i/M \in (0, 1]$;
　　$\Psi_i = e_i/M \in (0, 1]$;
　　$\varphi_i = t_i + t_i / T_i \in (0, 1]$;
　　$\rho_i = \begin{cases} j, j = c_j \\ j - 1, j > c_j \end{cases}$ ;
　　$\vec{W} = [0.1, 0.3, 0.3, 0.2, 0.1]$;
　　$R_i = 0.1\delta_i + 0.3\tau_i + 0.3\Psi_i + 0.2\varphi_i + 0.1\rho_i$;
　　$Y(t) = [(R_{i,n}/R_{i,t})^{1/t-1} - 1] * 100\%$;
　　update($\mathbb{Z}_{Old}$);
**end**
　$\mathbb{Z}_{New} = \mathbb{Z}_{Old}$;
　EBRC($\mathbb{Z}_{New}$);
　Return $\mathbb{Z}_{New}$;

ALGORITHM 1: Node evaluation.

the corresponding relationship between node permissions and their reputation value and reputation growth rate are shown in Table 3.

Based on Table 3, the verifiable random function (VRF) [36, 37] is used to randomly sample among the nodes with the election authority of candidate consensus nodes to select the consensus node-set. According to their reputation value ranking, the consensus nodes are divided into consensus nodes and candidate consensus nodes. The characteristics of the VRF random function are the following: for a specific random number seed input $S$ and the inputter's private key $SK$, the VRF function will output a random number L and Proof, and the verifier verifies whether the random number is correct and reliable through the random output number $L$, Proof and input seed $S$. Finally, we define a threshold $\omega$ for the random number $L$ output by the VRF and stipulate that nodes with the random number $L$ greater than or equal to the threshold can be a member of the consensus node-set. The definition of the threshold is mainly used to adjust the difficulty of node election.

Therefore, the detailed flow of the VRF election algorithm in this paper is as follows:

　(i) Based on the previously created block hash value, the current consensus master generates the seed $S$ of a random number for each epoch using a PRNG (pseudorandom number generator) in the last round of the consensus process. The current consensus master node will then add $S$ to the blockchain, and nodes across the network will see the seed random number

　(ii) Node $A$ determines whether it belongs to the consensus node-set, whether the random number $L$ calculated by $S$ using the VRF verification function is less than the threshold $\omega$. If that is the case, it broadcasts a connection request and waits for the remaining

TABLE 3: Election authority classification.

| Reputation ranking | Reputation growth rate ranking | Master node | Consensus node | Candidate consensus node |
|---|---|---|---|---|
| Top 25% | Top 25% | ✓ | ✓ | ✓ |
| Top 50% | Top 50% | ✓ | ✓ | ✗ |
| Top 85% | Top 85% | ✓ | ✗ | ✗ |
| Bottom 15% | Bottom 15% | ✗ | ✗ | ✗ |

consensus nodes to initiate connection requests within $\Delta h$. The formula is as follows:

$$L = \text{VRF}(SK, S),$$
$$\text{Proof} = \text{VRF\_Proof}(SK, S) \tag{4}$$

(iii) Any other node $B$ that receives a connection request can prove whether Proof is valid based on that node or not. If Proof is valid, continue; otherwise, abort. The formula is as follows:

$$\frac{L}{2^{256}} \in [0, 1] \leq \omega,$$
$$L = \text{VRF\_P2H}(\text{Proof}) \tag{5}$$

(iv) Node $B$ can then verify the public key $PK$, random number seed $S$, and Proof of node $A$. If the verification fails, it can report to the whole network and record the node's behavior in the behavior record table. If the verification is successful and node $B$ is a member of the consensus node-set, a connection request can be sent to node $A$ to complete the connection. The corresponding formula is as follows:

$$\text{True/False} = \text{VRF\_Verify}(PK, S, \text{Proof}) \tag{6}$$

(v) After each consensus node is connected, it ranks the reputation value according to the behavior record table. The top 50% nodes are consensus nodes, and the top 50% to 85% of nodes are candidate nodes. The consensus nodes will be replaced by candidates when they fail or exit

It should be noted that since $L/2^{256} \in [0, 1]$, also, the election algorithm here is probabilistic, so there may not be enough consensus nodes to conduct transactions. Moreover, after experimental testing, when the number of nodes is 10 and $\omega = 0.4$, the probability of a node not being selected is $P(0.4) = 0.01\%$. Therefore, we set $\omega = 0.4$ to ensure the ran-domness of node selection and minimize the probability of no selected node. The process is shown in Algorithm 2.

*2.4. EBRC Consensus Process.* This paper proposes EBRC to solve the consensus algorithm's scalability issues mentioned in the first section. The consensus process is shown in Figure 3.

In Figure 3, P1 is the master node and R3 is a malicious slave node; the consensus process is as follows:

(i) After the consensus node is selected, the current view $v$ is initialized, and the consensus master node $p$ of the current view is selected according to the master node election algorithm:

$$p = (h + v) \bmod (3f + 1), \tag{7}$$

where $h$ is the current block height and $f$ is the maximum number of nodes allowed to fail. A block is created during each round cycle, the view is numbered $v + 1$, and the primary node is replaced

(ii) When the transaction's initiator initiates the transaction, it signs the transaction with the private key and then broadcasts it to the entire network. The message format is

$$<<\text{Crequest}, t, d, m(d), c >, \text{Sig}_c >, \tag{8}$$

where $t$ is the timestamp, $d$ is the transaction data, $m(d)$ is the summary of the transaction data, $c$ is the client identification, and $\text{Sig}_c$ is the client signature

(iii) After the node receives the transaction, it can forward it if it is not a consensus node. If it is a consensus node, the legality of the transaction needs to be verified. If the transaction is valid, the node is ready to send a prepare message. If the transaction is illegal, it is directly discarded. When the master node sends the transaction information, the view replacement protocol can be quickly initiated without verifying the content of the message

(iv) After $\Delta t$ time ($\Delta t$ is the system built-in waiting time), the master node processes the request message to determine whether the message being processed is legitimate or not. If so, the master node $p$ broadcasts the consensus message to the slave node in the following form:

$$<<\text{Prepare}, h, v, t, d, m(d) >, \text{Sig}_p >, \tag{9}$$

where $h$ is the current block height + 1, $v$ is the view number, and $\text{Sig}_p$ is the signature of the primary node $p$

(v) After receiving a ready message from the master node, the message is verified to determine whether

**Input:** Behavior record table $\mathbb{Z}_{New}$. Number of consensus node-set $M$
**Output:** Consensus node-set $CON_S$, Candidate consensus node-set $CAN_S$, Consensus node $CON$
**While** $\mathbb{Z}_{New}$ **do**
    $Read(\mathbb{Z}_{New})$;
    $Sort(\mathbb{Z}_{New}, R)$;
**end**
$S = PRNG(previous_{hash})$;
$L = VRF(SK, S)$;
$Proof = VRFProof(SK, S)$;
**if** $L/2^{256} \leq \omega$ **then**
    $Broadcast - connectMessage(PK, Proof, Signature)$;
**else end**
**if** current.validator == consensus node and message.type == connect and node $i$ ranks in the top 85% in the reputation and reputationRate collections **then**
    $VRF_{verify}(message.PK, S, message.Proof)$;
    **if** $VRF_{verify}$ **then**
        $Add(CON_S, i)$;
        **if** $Count(CON_S \geq M)$ **then**
            $Sort(CON_S, R)$;
            $CON = top\ 85\%nodes$;
            $CAN_S = bottom\ 15\%nodes$ ;
        **else end**
    **else end**
**else end**
**return** $CON_S$, $CAN_S$, $CON$

ALGORITHM 2: Node election.

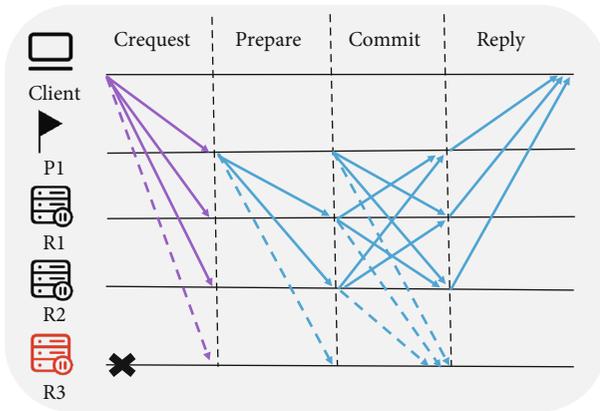

FIGURE 3: Efficient Byzantine consensus mechanism based on reputation (EBRC). The process consists of two phases: prepare and commit.

the master node is malevolent or malfunctioning. If so, a view conversion request is initiated, and the node identity is recorded in the node behavior table. Otherwise, the consensus node broadcasts a confirmation message to the master node. The message format is

$$<<Commit, v, t, m(d), sn, valid/invalid> , Sig_i> . \quad (10)$$

Among them, valid/invalid represents the identity of node $i$ to determine whether the message is valid, and $Sig_i$ is the signature of primary node $i$.

(vi) When all consensus nodes receive $2f + 1$ identical acknowledgment messages, a consensus is reached, and the client's request will be executed, the client will be replied to, and the block will be written, with a view number of $v + 1$. Otherwise, the view switching protocol is followed, and the node identity is reported in the node behavior table. The reply message is in the following format:

$$<<Reply, c, t, m(d), n, valid/invalid> , Sig_i> , \quad (11)$$

where $n$ is the number of consensus nodes, and the rest of the symbolic interpretation is as previously defined. The consensus protocol proposed in this paper significantly reduces the communication complexity in the PBFT consensus process. It should be noted that this modification is based on the reputation value and reputation growth rate of the node. The EBRC algorithm can only achieve a consensus when the node's reliability is guaranteed.

*2.5. Dynamic Joining and Exiting of Nodes.* DJEP (Dynamic Joining and Leaving Protocol) is intended to ensure that the system remains unaffected when nodes leave the consensus. If the system is not affected, the condition $(n - 1)/3$ can still be satisfied after the exit of the node or join.

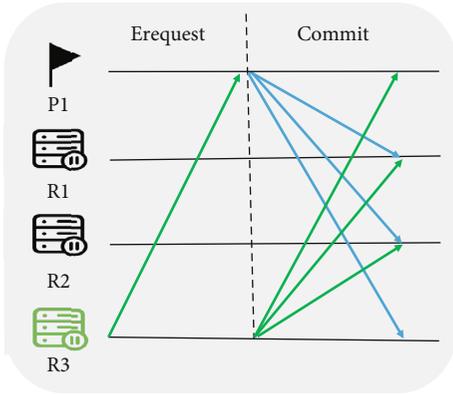

Figure 4: Dynamic exit of nodes. The process consists of two phases: Erequest and commit.

### 2.5.1. Dynamic Exit.
The dynamic exit mechanism for the node is shown in Figure 4 and consists mainly of the following processes:

(i) The exiting node sends an exit request to the master node in the form of

$$<<\text{ERequest}, h>, \text{Sig}_i>, \quad (12)$$

where $h$ is the block height + 1, representing the node exit time (that is, when the block height is $h + 1$)

(ii) If the number of consensus node-set becomes less than $3f + 1$, after the node exits, we then initiate the node dynamic joining phase. Otherwise, the master node and the node to be exited send exit confirmation messages with both parties' signatures to other slave nodes. The format of the confirmation message is as follows:

$$<<\text{Commit}, i, h> \text{Sig}_i/\text{Sig}_p>, \quad (13)$$

where $i$ is the number of nodes $i$ to exit, and this number is unique

### 2.5.2. Dynamic Join.
The dynamic join mechanism for nodes is shown in Figure 5 and consists of the following procedures:

(i) The master node sends a consensus node replacement message to the candidate consensus node. The candidate consensus nodes are ranked according to their reputation value, and the highest-ranked candidate node becomes the new consensus node. The replacement message format is as follows:

$$<<\text{Change}, i, h>, \text{Sig}_p>. \quad (14)$$

The symbolic meaning in the message is as described above.

(ii) The new consensus node sends a join request message to the master node and other slave nodes in the form of

$$<<\text{Urequest}, i, R, h>, \text{Sig}_i>, \quad (15)$$

where $R$ is the reputation value of node $i$ to be added. The consensus node then views the reputation of the candidate node and verifies the request

(iii) After the master node and the consensus node verify and confirm this message, they broadcast the "node joining confirmation message" to the consensus network so that the consensus node and the new consensus node can confirm each other. The format of the confirmation message is

$$<<\text{Commit}, i, h>, \text{Sig}_i> \quad (16)$$

## 3. Experimental Results

To evaluate our design, we use Hyperledger Sawtooth to develop a prototype of a blockchain system with EBRC suitable for IoT networks as a consensus protocol. To make a comparison with PBFT, we conduct experiments of original PBFT and EBRC. In this section, we present experimental results of system performance and network overhead.

### 3.1. Experiment Settings.
In order to verify the applicability of the EBRC consensus algorithm, we designed different experiment schemes and performed experiments on our cluster server machines; each of them contains a two-core Intel Core i7 2.2 GHz CPU with 16 GB DRAM and 256 GB SSD and Ubuntu 16.04 as an operating system. We installed Sawtooth PBFT 1.0 with EBRC to create simulation networks that contain the primary node, the secondary node, and the validator node. The initial consensus committee consists of 4 IoT devices, and the maximum number is set as 40. The performance analysis mainly examines the following indicators: transaction delay, throughput, communication times, election fairness, and the impact of joining and exiting on consensus.

### 3.2. Transaction Latency Analysis.
Transaction latency refers to transaction submission to transaction confirmation, which is the benchmark for calculating the communication efficiency and run time of the blockchain network's consensus algorithm. Lower the delay, accelerate transaction confirmation, and reduce the probability of the blockchain bifurcation phenomenon. The calculation formula is

$$T_{\text{delay}} = T_{\text{complete}} - T_{\text{submit}}, \quad (17)$$

where $T_{\text{delay}}$ is the latency, $T_{\text{complete}}$ is the time when the block is confirmed to be completed, and $T_{\text{submit}}$ is the time when the transaction starts executing.

We tested the transaction latency time of PBFT and EBRC algorithm under each of the following nodes: 4, 7,

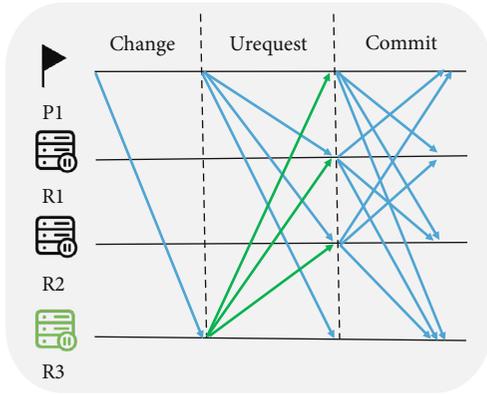

Figure 5: Dynamic join of nodes. The process consists of three phases: change, Urequest, and commit.

10, 13, 19, 25, 31, and 40, where Ns represents the number of nodes in the blockchain; 15 blocks with different numbers of transactions are tested. Every block was tested 15 times. And then, the latency of the EBRC algorithm and the PBFT algorithm was compared by taking the average latency of each block. As shown in Figure 6, the latency of the EBRC algorithm proposed in this article is significantly lower than that of PBFT, indicating that the node's consensus speed is faster than that of PBFT. It can also be seen that blocks with different numbers of transactions affect latency. The larger the block size, the more transactions it includes, and the longer it takes for network propagation and verification between nodes. Since network bandwidth is limited, the longer the broadcast confirmation period, the higher the latency. The results are shown in Figure 6.

It is easy to spot as the number of nodes $N$ increases, so does the PBFT consensus latency, which is caused by the $O(N^2)$ message complexity. As $N$ increases, the verification time increases quadratically. Although the EBRC algorithm latency also has a positive correlation with $N$, it generally grows slower than that of the PBFT algorithm. To sum up, this experimental result demonstrates a lower transaction latency of our proposed EBRC consensus.

### 3.3. Throughput Performance Analysis.

In a blockchain, transaction throughput per second (TPS) is an important index to measure the system's concurrency. The higher the throughput, the greater the consensus mechanism's efficiency, and the greater the ability to process transactions. The formula is

$$T_{tps} = \frac{T_{sum}}{\Delta t}, \quad (18)$$

where $T_{tps}$ stands for throughput, $T_{sum}$ stands for block time, and $\Delta t$ stands for the total number of transactions during that time. The throughput value is tested in this experiment by setting the upper limit of the transaction processed by each block. The client continuously sends transaction data to the database. 20 blocks were taken with different numbers of transactions for testing the throughput with and without the Byzantine nodes in the network.

Here, we did two sets of experiments:

(i) The client initiated 100 transactions, and we observed the transaction throughput performance of PBFT and EBRC algorithms under the different numbers of nodes $N$: 4, 7, 10, 13, 19, 25, 31, and 40. The result is shown in Figure 7.

As shown in the graph, the throughput of the two algorithms decreases as the nodes increase. However, the throughput of the EBRC algorithm is still significantly higher than that of the original PBFT algorithm.

(ii) We tested the transaction throughput between PBFT and EBRC algorithms when there are $(n - 1)/3$ error nodes and set the number of nodes in the entire network $N$ as 4, 7, 10, 13, 19, 25, and 31, respectively. In the experiment, the message received by the node is not forwarded to simulate the behavior of the malicious node. That is, the malicious node receives the message, but it does not feed back to the master node or other slave nodes. The result is shown in Figure 8.

The test observed that throughput is linked to the total number of transactions in a block, the number of nodes joining the consensus, and the number of malicious nodes in the network: when a network has fewer consensus nodes and no or few malicious nodes, the number of transactions that can be processed per unit time increases, and the throughput increases. As there are more malicious nodes in the network, the thread blocks and the throughput of PBFT decreases, but the throughput of EBRC remains higher than the original PBFT since the EBRC reputation consensus algorithm efficiently replaces malicious nodes with candidate consensus nodes and keeps the network running without affecting network throughput.

### 3.4. Communication Time Analysis.

The EBRC algorithm's communication times and the original PBFT algorithm in the consensus process were experimentally compared. The number of communications is the number of messages sent between nodes to complete the consensus, directly related to the computing resources required to deploy the algorithm. Specifically, under different numbers of nodes, and the Byzantine node is still $(n - 1)/3$, the measured number of node communications is shown in Figure 9.

It can be seen from Figure 9 that the communication time of the PBFT algorithm and EBRC algorithm has a positive correlation with the number of nodes. However, the speed increment of the EBRC algorithm is relatively slow. Reduced communication can effectively save computational resources, minimize server load, and increase the number of consensus nodes that the algorithm can accommodate.

### 3.5. Election Performance Analysis.

Since the EBRC algorithm adds a reputation-based voting mechanism, the voting performance's experimental analysis is carried out. The experiment is divided into two parts:

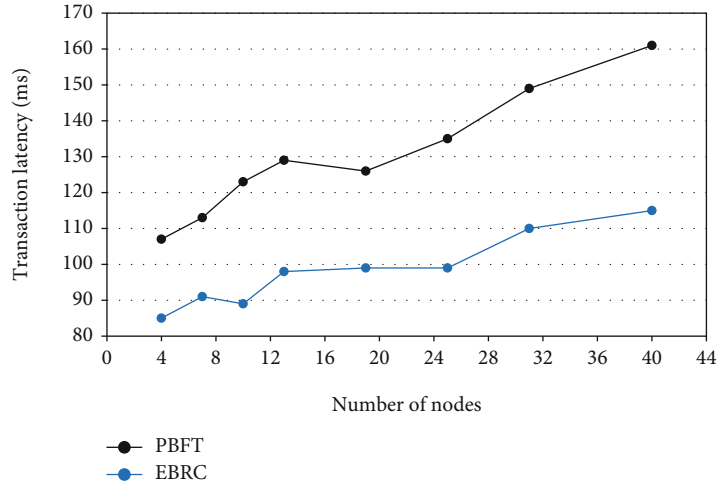

FIGURE 6: The relationship between the latency of PBFT and EBRC and the number of nodes.

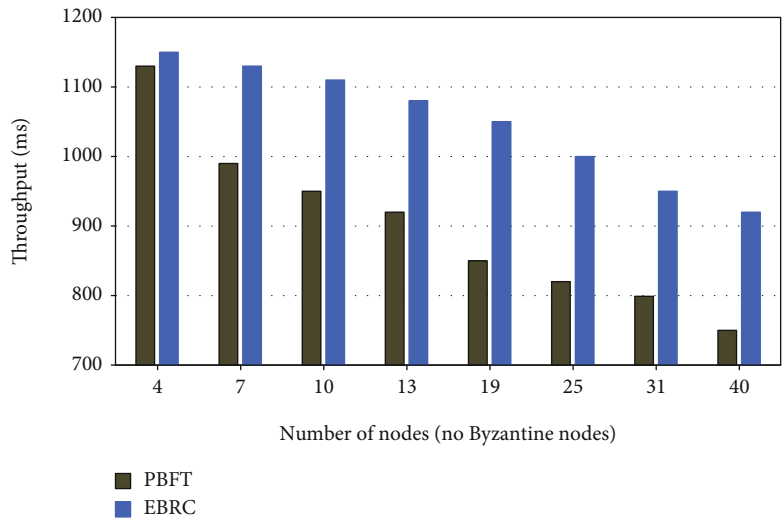

FIGURE 7: PBFT versus EBRC throughput with different numbers of nodes (no Byzantine nodes).

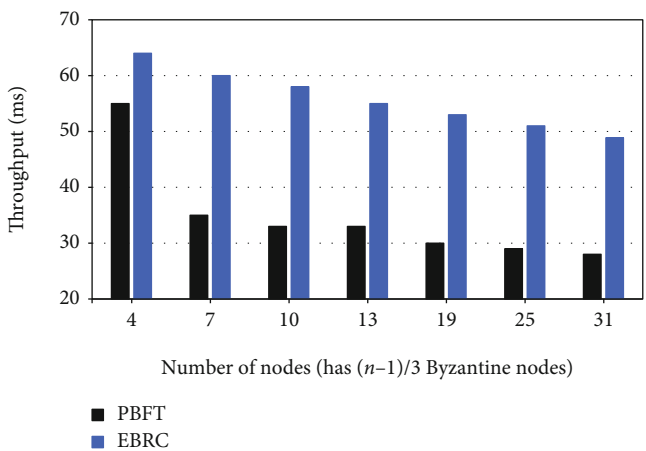

FIGURE 8: Comparison of throughput between PBFT and EBRC under different numbers of nodes (the system has $(n-1)/3$ Byzantine nodes).

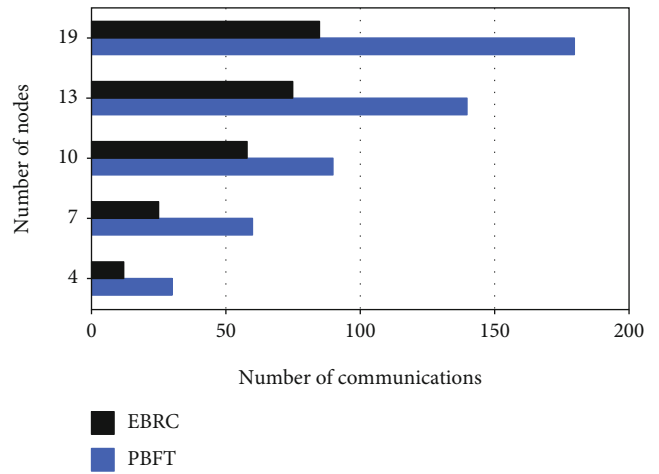

FIGURE 9: Comparison of communication times between nodes.

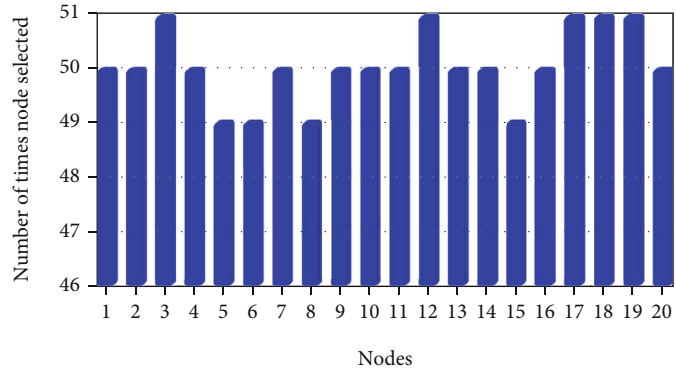

FIGURE 10: Electoral performance analysis. In order to determine the fairness of the election algorithm, the number of nodes with similar reputations is compared with the number of nodes being elected.

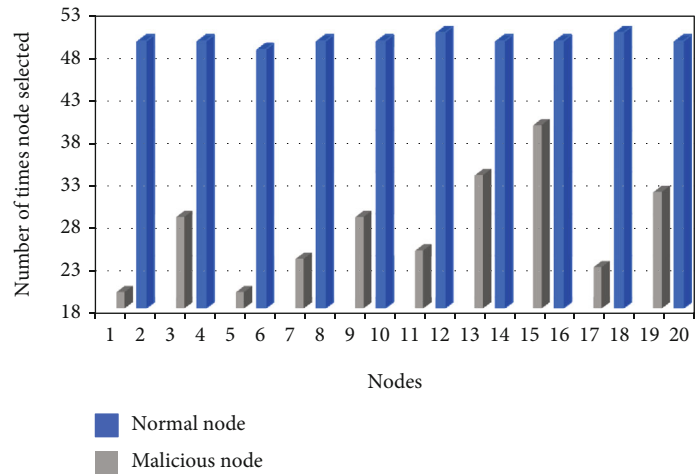

FIGURE 11: Electoral performance analysis. The number of normal nodes and malicious nodes being elected is compared to determine the practicability of the election algorithm.

TABLE 4: Comparison of relevant consensus algorithms.

| Consensus algorithms | Node reliability evaluation | Node election fairness | Substantial punishment for nodes | Communication complexity | Suitable for dynamic networks |
|---|---|---|---|---|---|
| PBFT | × | × | × | $O(n^2)$ | × |
| DBFT | × | × | × | $O(n^2)$ | ✓ |
| EPBFT | × | ✓ | × | $O(n)$ or $O(n^2)$ | ✓ |
| Trust-PBFT | ✓ | × | × | $O(n^2)$ | × |
| T-PBFT | ✓ | × | × | $O(n^2)$ | × |
| Tu | ✓ | × | × | $O(m^2), m < n$ | × |
| EBRC | ✓ | ✓ | ✓ | $O(m^2), m < n$ | ✓ |

(i) Firstly, we create 20 nodes and set all the nodes' reputations to a default value of 0.5. We ran 1,000 experiments to give all the nodes enough time to evaluate. We tested the number of times the node was elected as a consensus node (including candidate consensus nodes), as shown in Figure 10. The number of times that a node is elected is an even distribution, ranges from 48 to 51, proving that the election of nodes is fair.

(ii) Secondly, we adjusted their reputation value by increasing the error rate of nodes with odd ids. The

experimental results are shown in Figure 11. It can be seen that when the error rate increases, the number of times that the node is elected as a consensus node is significantly reduced, which is two times less than the number of ordinary node elections, which shows that if the node conducts suspicious actions, our algorithm can reduce the node's reputation value and the probability of a successful election.

It is also observed that as the error rate increases, the number of times a node is elected as a consensus node decreases fourfold compared to a regular node, demonstrating that if the node does not follow the protocol, the decreased credibility value will impact the election's chances of success. It also takes a long time for nodes to recover from disruptive actions.

### 3.6. Dynamic Join and Exit Performance Analysis.

This section will verify the dynamic exit and join mechanism DJEP proposed in this article. Since this mechanism is a special stage in the consensus process, it is necessary to test for the delays that increase during the consensus phase after triggering this mechanism. We separately tested the RC algorithm's transaction delay when the number of nodes is 25, the client initiates 15 transactions, and the node initiates join and exit requests. The experiments are as follows:

  (i) When the node exit, the total number of consensus nodes meets the condition of $n \geq (n-1)/3$, which only triggers the node exit mechanism

  (ii) After the node exit, the total number of consensus nodes does not meet the condition of $n \geq (n-1)/3$, and the node exit and join mechanism is triggered

Experimental findings show that in experiment (1), the average consensus delay is 100 ms, which is 1 ms longer than the standard consensus phase delay, where the increased delay is almost negligible. There are fewer communications between nodes in the dynamic exit mechanism of nodes, so the consensus' impact is negligible. In experiment (2), the average consensus delay is 127 ms, which is 28 ms longer than the normal consensus phase delay. In this case, there are more communications between nodes in the dynamic exit and join mechanism of nodes, so the increased delay is more. However, this delay is also less than the delay of the PBFT algorithm under the same conditions.

Through the above experimental analysis, we have concluded that comparing with the PBFT algorithm, the EBRC algorithm improves the reliability and safety of the nodes in the system and the dynamics and scalability of the system. Besides, the algorithm also reduces system latency and improves system throughput.

### 3.7. Comparison of Relevant Consensus Algorithms.

As shown in Table 4, we compare PBFT, DBFT, EPBFT, Trust-PBFT, T-PBFT, Tu, and EBRC.

We can see that EBRC has the following advantages compared with other consensus algorithms.

  (1) EBRC has improved consensus node reliability evaluation, making it ideal for IoT networks with significant performance gaps and maintaining consensus node reliability and robustness

  (2) EBRC adds a random selection of consensus nodes using VRF based on reputation, making it suitable for IoT networks with a large number of nodes. Moreover, the VRF-based election method is more secure and fair

  (3) EBRC can identify and punish malicious nodes. Identifying and punishing these malicious nodes will improve the reliability of the system, reduce the probability of nodes doing evil, and increase the speed of reaching consensus

  (4) EBRC reduces the communication overhead in a stable network. In this network, the communication complexity required to reach a consensus is reduced from $O(n^2)$ to $O(n)$

  (5) EBRC also enhances the dynamics of the IoT blockchain network through the dynamic joining and exit mechanism of nodes, making it more suitable for the real Internet of Things blockchain network environment

## 4. Conclusion

With the increasing popularity of IoT blockchain applications, the performance and scalability of IoT blockchain systems have become increasingly critical. To solve the poor scalability and high overhead in existing IoT blockchain applications, we propose EBRC, a scalable and reputation-based consensus protocol. The proposed EBRC achieves high consensus efficiency, low network overhead, and high scalability by reputation-based node election and dynamic joining and exiting protocol of nodes. By integrating the VRF random election algorithm, we randomly selected the high reputation node to join the consensus. We authenticated IoT devices in the node sets to take the consensus role. Finally, extensive experiments were conducted to indicate the superior performance of EBRC over the traditional PBFT consensus mechanism, indicating that our proposed algorithm can provide an effective solution for the construction of the Internet of Things+blockchain+Internet court.

## Data Availability

The data in this paper are from the original experiment.

## Conflicts of Interest

The authors declare that they have no conflicts of interest.

## Acknowledgments

We sincerely thank Ms. Yike Zhang for her help in this article. This work is supported by the National Key Research and Development Program of China under Grant 2018YFC0831305.